\documentclass[a4paper, superscriptaddress, reprint]{revtex4-1}

\usepackage[utf8x]{inputenc} 

\usepackage[english]{babel}

\usepackage{times}

\usepackage{lmodern}
\usepackage{textcomp}
\usepackage[T1]{fontenc}

\usepackage{graphicx}
\graphicspath{{./pictures/}}
\usepackage{floatrow}
\newlength{\figwidth}
\usepackage{booktabs}

\usepackage{amsmath}

\usepackage[sfdefault=cmbr]{isomath}
\usepackage{siunitx}

\usepackage{braket}

\usepackage[breaklinks = true]{hyperref}
\usepackage[all]{hypcap}
\usepackage{url}

\usepackage{cleveref}

\usepackage{tikz}
\usetikzlibrary{external}

\usepackage[version=3]{mhchem}

\providecommand*{\ch}[1]{\ce{#1}}
\providecommand*{\ped}[1]{\ensuremath{_\textup{#1}}}		

\hypersetup{
pdfauthor = {Marco Pizzocaro et al.},
colorlinks,%
citecolor=black,%
filecolor=black,%
linkcolor=black,%
urlcolor=blue
}

\begin{document}
\title{Absolute frequency measurement of the \texorpdfstring{\ch{^1S_0} -- \ch{^3P_0} transition~of~\ch{^{171}Yb}}{1S0 -- 3P0 transition of 171Yb}}

\begin{abstract}
We report the absolute frequency measurement of the unperturbed transition  \ch{^1S_0} -- \ch{^3P_0} at \SI{578}{nm} in  \ch{^{171}Yb} realized in an optical lattice frequency standard relative to a cryogenic caesium fountain.
The measurement result is \SI{518295836590863.59(31)}{Hz}  with a relative standard uncertainty of \num{5.9e-16}.
This value is in agreement with the ytterbium frequency recommended as a secondary representation of the second in the International System of Units.
\end{abstract}


\author{Marco~Pizzocaro}
\affiliation{Istituto Nazionale di Ricerca Metrologica (INRIM), Physical Metrology Division, Strada delle Cacce 91, 10135 Torino, Italy}

\author{Pierre~Thoumany}
\affiliation{Istituto Nazionale di Ricerca Metrologica (INRIM), Physical Metrology Division, Strada delle Cacce 91, 10135 Torino, Italy}

\author{Benjamin~Rauf}
\affiliation{Istituto Nazionale di Ricerca Metrologica (INRIM), Physical Metrology Division, Strada delle Cacce 91, 10135 Torino, Italy}
\affiliation{Politecnico di Torino, Dipartimento di Elettronica e Telecomunicazioni, Corso duca degli Abruzzi 24, 10129 Torino, Italy}

\author{Filippo~Bregolin}
\affiliation{Istituto Nazionale di Ricerca Metrologica (INRIM), Physical Metrology Division, Strada delle Cacce 91, 10135 Torino, Italy}
\affiliation{Politecnico di Torino, Dipartimento di Elettronica e Telecomunicazioni, Corso duca degli Abruzzi 24, 10129 Torino, Italy}

\author{Gianmaria~Milani}
\affiliation{Istituto Nazionale di Ricerca Metrologica (INRIM), Physical Metrology Division, Strada delle Cacce 91, 10135 Torino, Italy}
\affiliation{Politecnico di Torino, Dipartimento di Elettronica e Telecomunicazioni, Corso duca degli Abruzzi 24, 10129 Torino, Italy}

\author{Cecilia~Clivati}
\affiliation{Istituto Nazionale di Ricerca Metrologica (INRIM), Physical Metrology Division, Strada delle Cacce 91, 10135 Torino, Italy}

\author{Giovanni~A.~Costanzo}
\affiliation{Istituto Nazionale di Ricerca Metrologica (INRIM), Physical Metrology Division, Strada delle Cacce 91, 10135 Torino, Italy}
\affiliation{Politecnico di Torino, Dipartimento di Elettronica e Telecomunicazioni, Corso duca degli Abruzzi 24, 10129 Torino, Italy}

\author{Filippo~Levi}
\affiliation{Istituto Nazionale di Ricerca Metrologica (INRIM), Physical Metrology Division, Strada delle Cacce 91, 10135 Torino, Italy}

\author{Davide~Calonico}
\email[Author to whom correspondence should be addressed. Electronic mail: ]{d.calonico@inrim.it}
\affiliation{Istituto Nazionale di Ricerca Metrologica (INRIM), Physical Metrology Division, Strada delle Cacce 91, 10135 Torino, Italy}



\maketitle


\section{Introduction}
Time and frequency are the physical quantities that can be measured with the lowest uncertainty so far and atomic frequency standards based on the microwave hyperfine transition of the caesium ground state realize the second in the International System of Units (SI).
The fractional uncertainties of the today's best Cs fountains are in the low \num{e-16} region \cite{Levi2014,Heavner2014,Guena2012} while optical frequency standards are now surpassing the caesium standards both in terms of accuracy and stability. 
Optical frequency standards using various atomic species and reference transitions \cite{Ludlow2015,Ushijima2015,Yamanaka2015,Targat2013,Lemke2009,Huntemann2016,Madej2012,Rosenband2008,Lorini2008,Huang2016} are being developed around the world also motivated by the prospect of a redefinition of the SI second based on an optical transition \cite{Gill2016}.
Optical frequency standards are of interest as highly sensitive probes for fundamental physics such as relativistic geodesy \cite{Bjerhammar1985,Lisdat2016,Takano2016}, tests of relativity \cite{Chou2010}, quantum simulations \cite{Inguscio2013}, and  searches for the variation of fundamental constants \cite{Huntemann2016,Godun2014,Blatt2008}.

In optical lattice clocks thousands of neutral atoms are trapped in an optical potential at the light-induced Stark shift cancellation (magic) wavelength \cite{Katori2003}. 
Optical lattice frequency standards have now demonstrated systematic fractional uncertainties at the \num{e-17}--\num{e-18} level \cite{Ushijima2015} and also an unprecedented frequency stability \cite{Hinkley2013} opening the path for frequency comparisons beyond the uncertainty of the realization of the SI units in hundreds of seconds \cite{Nemitz2016}.
In particular, different groups developed ytterbium optical lattice frequency standards worldwide \cite{Nemitz2016, Hinkley2013, Akamatsu2014, Park2013} and the spin and angular-momentum forbidden transition \ch{^1S_0} -- \ch{^3P_0} of ytterbium~171  at \SI{578}{nm}  is recommended as a secondary representation of the SI second  by the International Committee for Weights and Measures (CIPM).
Absolute frequency measurement of this transition has been performed relative to the realization of the second with caesium standards at the National Institute of Standards and Technology (NIST)\cite{Lemke2009}, at the National Metrology Institute of Japan (NMIJ) \cite{Kohno2009,Yasuda2012}, and at the Korea Research Institute of Standards and Science (KRISS) \cite{Park2013}.
Optical frequency ratios between ytterbium and strontium lattice standards were performed at NMIJ \cite{Akamatsu2014} and  RIKEN \cite{Takamoto2015, Nemitz2016}.



We designed and operated an optical lattice frequency standard based on neutral ytterbium atoms.
We report the frequency measurement of the unperturbed clock transition \ch{^1S_0} -- \ch{^3P_0} of \ch{^{171}Yb} against the cryogenic Cs fountain ITCsF2  \cite{Levi2014} with a total fractional uncertainty of \num{5.9e-16}. The systematic uncertainty of the Yb optical lattice frequency standard is below that of the Cs fountain as well as the statistic uncertainty.
Absolute frequency measurements of optical frequency standards are important to show confidence in their reproducibility and are at the base of a possible redefinition of the SI second
based on an optical transition.

The paper is organized as follows: in Sec. 2 we describe the realization of the Yb lattice frequency standard. In Sec. 3 we report on the study of systematic frequency shifts and obtain the uncertainty budget for the Yb frequency standard. This is followed in Sec. 4 by a description of the absolute frequency
measurement. We  finally report and discuss the results.

\begin{figure}
\includegraphics[width=\figwidth]{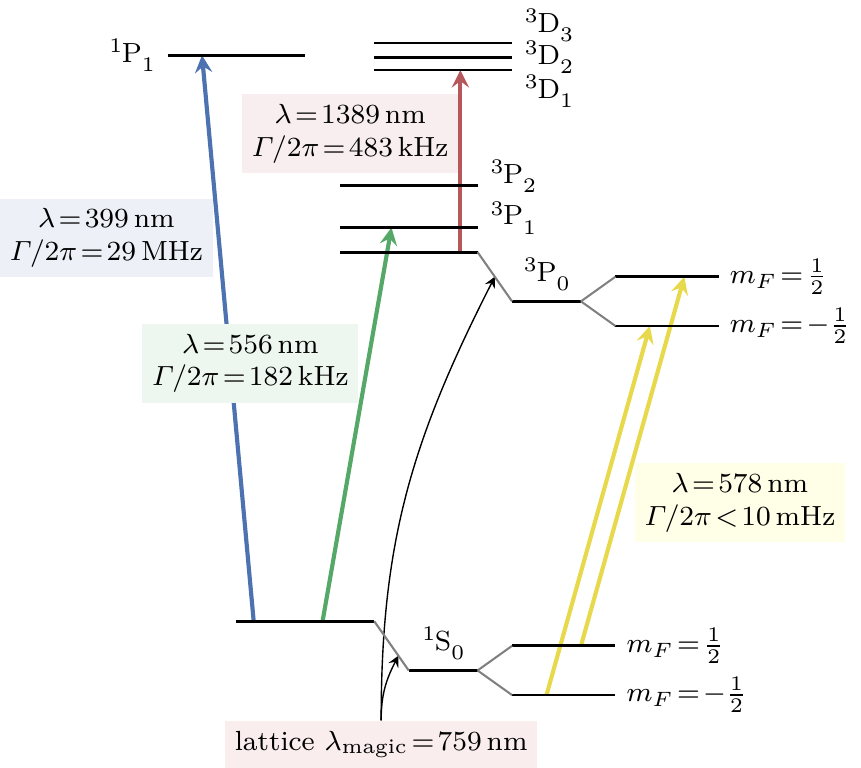}
\caption{Partial term diagram for $\ch{^{171}Yb}$ with transition wavelengths and natural linewidths relevant for the operation of the standard. The lattice shift and the Zeeman components are shown for the clock states $\ch{^1S_0}$ and $\ch{^3P_0}$.
}
\label{fig:levels}
\end{figure}

\section{Ytterbium lattice frequency standard}

\Cref{fig:levels} shows the energy levels relevant for the cooling, trapping, and state manipulation of ultracold  $\ch{^{171}Yb}$ atoms and for the operation of the frequency standard.
$\ch{^{171}Yb}$ is fermionic with nuclear spin $I=1/2$.

The clock transition \ch{^1S_0} -- \ch{^3P_0} is at \SI{578}{nm} and has a natural linewidth <\SI{10}{mHz} \cite{Porsev2004}.
The magic wavelength  where the light-induced scalar Stark shift cancels for the transition is \SI{759}{nm}.
A bias magnetic field splits the clock transition  in two resonances from each $m_F\!=\!\pm1/2$ state.
We interrogate with a linearly-polarized probe the two $\pi$-transitions $m_F \!=\! 1/2$ -- $m_F \!=\! 1/2$ and $m_F \!=\! -1/2$ -- $m_F \!=\! -1/2$ of which we take the average.
By averaging over the two $\pi$-transitions we cancel the first-order Zeeman shift and the vector components of the light shift induced by the lattice \cite{Lemke2009}.
The tensor components of the light shift is zero for spin $1/2$ atoms.
Furthermore, the atomic ensemble can be prepared in a single spin state enhancing the spectroscopic signal and reducing the shift due to cold collisions \cite{Ludlow2011}.


Ytterbium atoms can be cooled to microkelvin temperatures in a two stages magneto-optical-trap (MOT) exploiting the strong \ch{^1S_0} -- \ch{^1P_1} transition  at \SI{399}{nm} (linewidth \SI{29}{MHz}) and the weaker \ch{^1S_0} -- \ch{^3P_1} transition  at 556 nm (linewidth 182 kHz).
Then the atoms are loaded in a one-dimensional optical lattice engineered by a retro-reflected laser beam at the magic wavelength \SI{759}{nm}.
The atoms are  prepared in a single spin state using an optical pumping scheme on the \ch{^1S_0} -- \ch{^3P_1} transition.
We probe the clock transition using Rabi spectroscopy with a \SI{60}{ms} pulse at \SI{578}{nm}.
A repumper at \SI{1389}{nm} resonant with the \ch{^3P_0} -- \ch{^3D_1} transition can be used to pump atoms out of the excited state~\ch{^3P_0}  to the ground state~\ch{^1S_0}.
Finally, a normalized state detection is performed measuring the \SI{399}{nm} fluorescence with a photomultiplier tube (PMT).
The total duration of a single spectroscopy sequence is \SI{310}{ms}.   

\begin{figure*}
\centerline{\includegraphics[width=6in]{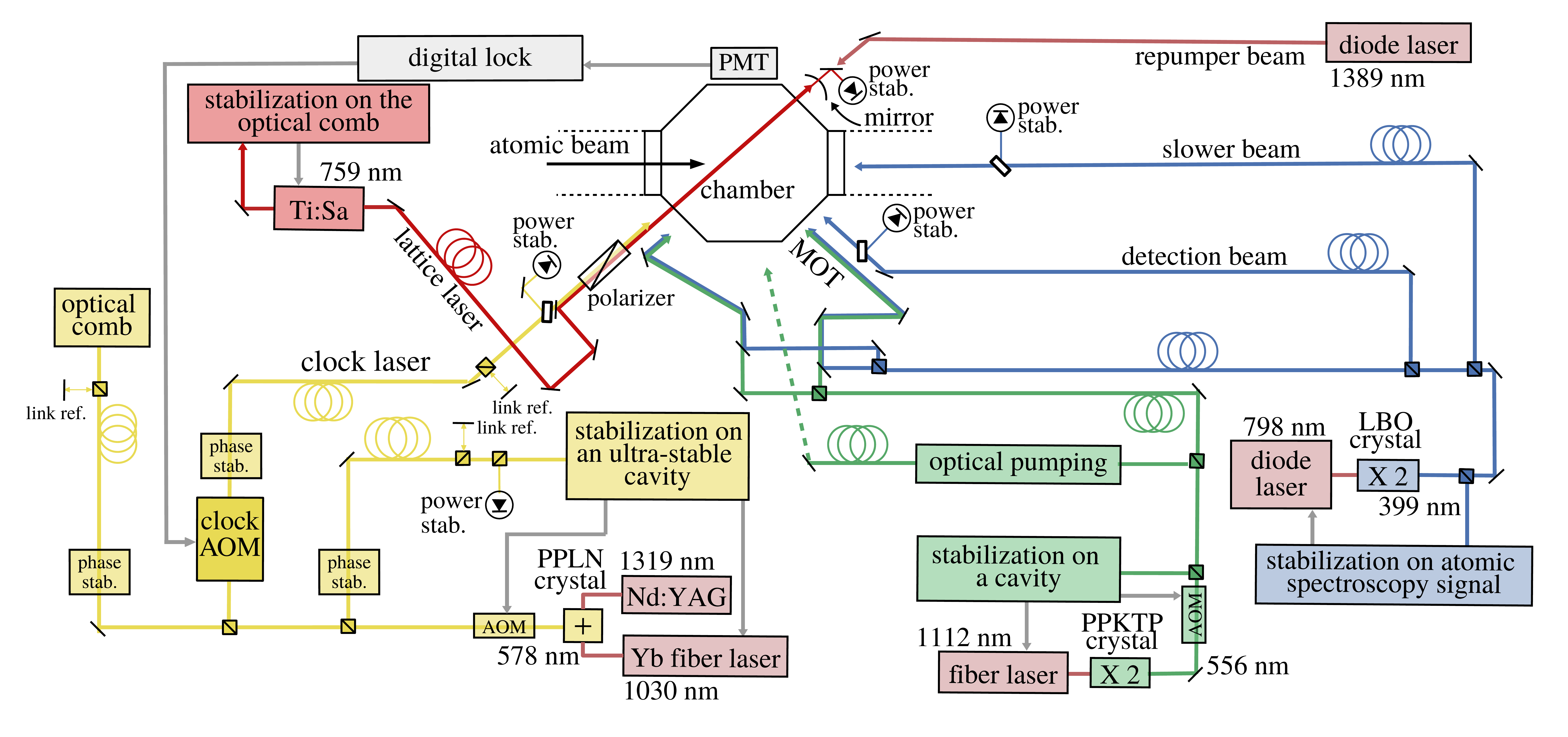}}
\caption{Experimental setup of the ytterbium optical lattice frequency standard showing the lasers generation and the delivery of the radiation to the trapping chamber.} \label{fig:setup}
\end{figure*}

\Cref{fig:setup} displays the experimental setup for atomic spectroscopy. 
The atomic source is an effusion oven at \SI{400}{\celsius} that produces a collimated atomic beam.
The distance between the atoms source and the trapping chamber is kept minimal (\SI{21}{cm}) to increase the atomic flux without implementing a Zeeman-slower.
The trapping region is a custom aluminum chamber designed for wide optical access with indium-sealed viewports.
Ten platinum resistance thermometers are distributed outside the trapping chamber for precise  evaluation of the black-body shift.
Standard stainless-steel fittings are used for the oven and for the main chamber pumping.
Two ion pumps and one non-evaporable getter pump maintains the ultra-high vacuum (pressure $<\SI{e-9}{mbar}$).
A pair of anti-Helmholtz coils generate the magnetic gradient for the MOT.
These are vertically arrayed outside the vacuum chamber and are water-cooled by a  chiller. Three pairs of Helmholtz coils are compensating the stray magnetic field. Furthermore, all lasers are spatially separated from the vacuum chamber and are delivered to the atoms by polarization-maintaining  optical fibers. 

\subsection{First stage MOT at \SI{399}{nm}}
Atoms are trapped from the atomic beam in the first stage MOT using the strong \ch{^1S_0}$(F=1/2)$ -- \ch{^1P_1}$(F=3/2)$ transition at \SI{399}{nm}.
The \SI{399}{nm} radiation is obtained by cavity-enhanced second harmonic generation (SHG) from a amplified, external-cavity diode laser at \SI{798}{nm} with an output power up to \SI{3}{W}.
The nonlinear medium is a lithium-triborate  (LBO) crystal \cite{Pizzocaro2014}. 
The diode laser is fiber coupled for spatial mode cleaning and we generate up to \SI{400}{mW} of light at \SI{399}{nm} with an input power of \SI{1}{W} into the enhancing cavity. The laser is locked to the \ch{^1S_0} -- \ch{^1P_1} transition through transverse  spectroscopy  of  an  auxiliary  ytterbium atomic  beam.

The six MOT beams add up to about \SI{30}{mW} of light  power with an $1/e^2$ radius of \SI{1}{cm} and a detuning of \SI{-20}{MHz}.
The magnetic field gradient is \SI{0.35}{T/m} along the z-axis. 
The atomic beam is slowed down by a counter-propagating \SI {399}{nm} light beam (slower beam) which has a power of about \SI{35}{mW} and a detuning from resonance of \SI{-360}{MHz}.
The leaking field of the MOT coils is exploited to slow the atoms, avoiding the use of an external spatially-varying magnetic field.
With the slower beam, we can capture up to \num{4e7} \ch{^{171}Yb} atoms.
Approximately \num{2e5} atoms are trapped and cooled down to about \SI{1}{mK} in \SI{50}{ms} during the clock cycle.

\subsection {Second stage MOT at \SI{556}{nm}}
The second  stage  MOT relies on the narrower \ch{^1S_0}$(F=1/2)$ -- \ch{^3P_1}$(F=3/2)$ transition at \SI{556}{nm}. 
The \SI{556}{nm} radiation is  generated by SHG of an amplified \SI{1112}{nm} Yb-doped fiber laser in single pass through periodically-poled  potassium titanyl phosphate (PPKTP) crystal.
About \SI{10}{mW} of \SI{556}{nm} radiation are obtained from \SI{1}{W} of infrared light.
The laser frequency is locked to a 10 cm stable cavity made of Corning  Ultra-Low Expansion glass  (ULE).  
The \SI{556}{nm} MOT beams have a total power of \SI{2}{mW}, a $1/e^2$ radius of \SI{0.5}{cm} and are kept on during the first MOT stage.
The second stage \SI{556}{nm} MOT is loaded from the  \SI{399}{nm} MOT with an efficiency up to the \SI{70}{\%} when the \SI{399}{nm} beams are turned off.
In \SI{60}{ms} we apply 3 substages at different frequency, intensity, and magnetic field gradient to maximize the fraction of atoms transferred into the lattice.
The first substage (\SI{30}{ms}) is designed for maximum atoms transfer between the two MOTs using  a  magnetic  field  gradient of \SI{0.13}{T/m} and maximum beam intensity.
The second substage (\SI{20}{ms}) decreases the atomic temperature down to \SI{10}{\micro K} moving the frequency closer to resonance, reducing the intensity 10 times,  and increasing the magnetic field gradient to \SI{0.2}{T/m}.  The  last  substage (\SI{10}{ms} and \SI{0.23}{T/m}) provides an enhanced transfer efficiency (typically \SI{7}{\percent}) of the atoms into the red-detuned optical lattice.

\subsection {Optical lattice at \SI{759}{nm}}
The optical lattice at the magic wavelength \SI{759}{nm} is provided by a titanium-sapphire laser with an initial output power of about \SI{2}{W}.
The optical lattice is formed by a  focused beam tilted from the horizontal plane by \SI{60}{mrad} and retro-reflected by a curved mirror.
The light is delivered using a dedicated fiber and is focused to a $1/e^2$ waist radius of \SI{45} {\micro m}. 
After the fiber the laser power incident on the atoms reaches \SI{1}{W}.
The operating depth of the lattice is  \SI{200}{$E\ped{r}$}  where $E\ped{r}=\hbar^2k\ped{L}^2/(2m)$ is the recoil energy from a lattice photon with wavenumber $k\ped{L}$, $\hbar$ is the reduced Planck constant, and $m$ the mass of the atom.
During a single cycle up to \num{1e4} atoms are trapped with a lifetime in the lattice of \SI{2.7(1)}{s}.

Control of the trapping conditions, thus of the light-induced frequency shifts on the clock transition, requires locking of the lattice frequency at the megahertz level and intensity stabilization.
The  titanium-sapphire laser has a linewidth below \SI{20}{kHz} and its frequency is measured with an optical frequency comb.
The slow drift of the laser is compensated by implementing a digital lock on the beat-note signal between the laser and the frequency comb.
The intensity and thus the lattice trap depth is stabilized by acting on the diffraction efficiency of an acousto-optic modulator (AOM) placed before the fiber. 
Finally linear polarization of the lattice light is achieved by a Glan-Thompson polarizer placed in front of the input window. 

\subsection {Single spin state preparation}
    
Spin-polarization of the ground state is achieved using optical pumping on the \ch{^1S_0}$(F=1/2)$ -- \ch{^3P_1}$(F=3/2)$ transition at 556 nm.
A linearly-polarized \SI{556}{nm} pulse and a bias magnetic field of  \SI{0.4}{mT}  are applied along the  vertical  axis for \SI{1}{ms}. 
Atoms in the lattice are prepared in either single spin ground state with \SI{98}{\percent} efficiency by choosing the \SI{556}{nm} pulse frequency (see \cref{fig:spinpol}).
After spin-polarization we wait \SI{30}{ms} to let magnetic transients decay.

\subsection{Clock transition spectroscopy}
The \SI{578}{nm} ultrastable laser 
is generated by sum-frequency-generation (SFG) in a waveguide periodically-poled lithium niobate  (PPLN) crystal combining a \SI{1319}{nm} neodymium-doped yttrium aluminum garnet (Nd:YAG) laser with an erbium fiber laser at \SI{1030}{nm} \cite{Pizzocaro2012}.
The total power obtained at \SI{578}{nm} is about \SI{7}{mW}.

The frequency of the \SI{578}{nm} laser is actively stabilized via the Pound-Drever-Hall (PDH) method on a \SI{10}{cm} ultrastable cavity made in ULE featuring fused-silica mirrors with ULE compensation rings.
The cavity is suspended horizontally on 4 points in a vibration-insensitive configuration and held in an antiacoustic chamber on top of a passive seismic-isolation platform.
A two-stage control using Peltier elements keeps the cavity at the temperature where the thermal expansion coefficient of ULE vanishes (about \SI{28}{\celsius} for our cavity).
The temperature servo is a digital implementation of the Active Disturbance Rejection Control (ADRC) technique \cite{Pizzocaro2013}. The laser is distributed to the ultrastable cavity, to the vacuum chamber for spectroscopy, and to the fiber-based optical comb by polarization-maintaining optical fibers. 
All fiber links are actively phase-stabilized using AOMs.
From measurements with similar cavities we expect this laser to have a flicker frequency instability less than \num{3e-15} at \SI{1}{s} of integration time \cite{Pizzocaro2012}.
The cavity drift is less than \SI{0.1}{Hz/s} and we use a feed-forward correction on a synthesizer to cancel it.

\begin{figure}
\includegraphics[width=\figwidth]{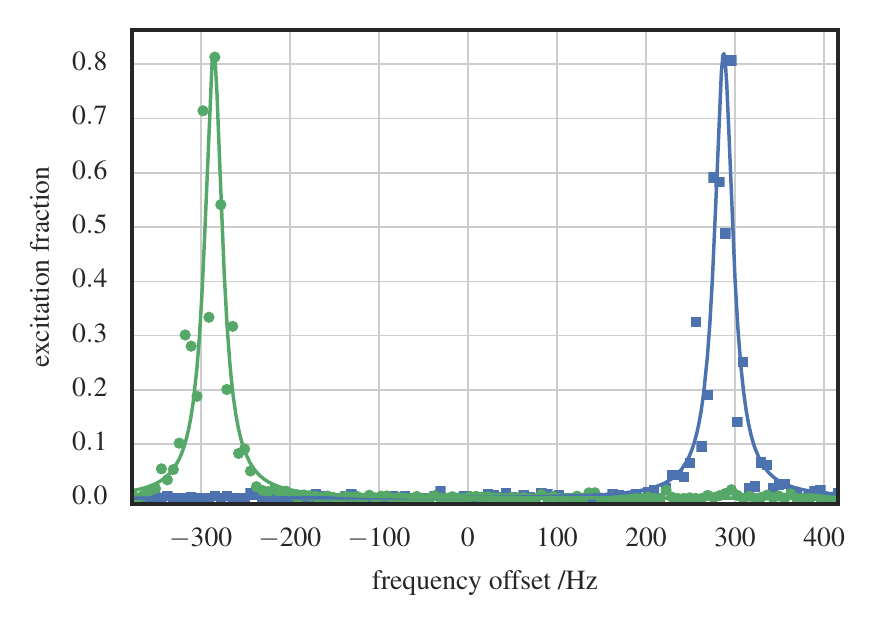}
\caption{Measured \ch{^1S_0} -- \ch{^3P_0} transition in presence of an external magnetic field of $B=$ \SI{0.13} {mT} and $\pi$-polarized light with the atoms polarized in the $m_F \!=\! +1/2$ state (green dots) or  $m_F \!=\! -1/2$ state (blue squares). Solid lines are lorentzian fits of the resonances. We observed a \SI{98}{\percent} single spin state purity.}
\label{fig:spinpol}
\end{figure}

Spectroscopy of the clock transition is achieved with a \SI{578}{nm} laser beam with a $1/e^2$ waist radius of \SI{200}{\micro m} collinear to the lattice with an incident power of \SI{70}{nW}.
The lattice retro-reflector is anti-reflection coated at \SI{578}{nm} with a reflectivity of $<\SI{1}{\percent}$ at this wavelength.
The probe linear polarization is aligned with the lattice polarization and to the vertical bias magnetic field using the same Glan-Thompson polarizer.
The frequency  of the probe light is controlled by a double-pass AOM placed before the fiber link to the atoms  (clock AOM in \cref{fig:setup}).
Probe intensity is actively stabilized.
During the interrogation a \SI{0.14}{mT} magnetic field is applied leading to a \SI{290}{Hz} Zeeman shift of the two $\pi$-transitions from the unperturbed one (\cref{fig:spinpol}).
Pulsing of the clock laser is achieved by detuning the applied frequency by \SI{1}{MHz}.
A $\pi$ pulse is applied for a Rabi time of \SI{60}{ms}.
Mechanical shutters stop all other laser beams during spectroscopy.

The resulting excitation fraction is then destructively detected and normalized applying three resonant pulses at \SI{399}{nm} and by collecting the atomic fluorescence with the PMT. 
The detection beam has a power of \SI{2.5}{mW}, a waist of \SI{0.5}{mm} and it strongly saturates the \ch{^1S_0} -- \ch{^1P_1} transition.
For each pulse the PMT signal is digitally integrated for \SI{2}{ms}.
The first pulse measures the population in the \ch{^1S_0} state.
The second one  reveals the background signal from scattered light and excited hot background gas atoms.
A third \SI{399}{nm} pulse then reveals the atoms that had been excited during the Rabi spectroscopy  after repumping the atom population of the \ch{^3P_0} state over the short-lived \ch{^3D_1} level back to the \ch{^1S_0}. 
For this task we use a commercial pigtail distributed feedback laser emitting at \SI{1389}{nm}.
The \SI{10}{mW} output beam of the repumper laser is focused on the atoms thus broadening the \ch{^3P_0} -- \ch{^3D_1} transition to \SI{300}{MHz}.
Active frequency stabilization is not required and the population transfer efficiency is \SI{90}{\%} after \SI{6}{ms}. 
We use the fluorescence from the three pulses to digitally calculate the normalized excitation fraction and the total number of atoms.

\begin{figure}
\includegraphics[width=\figwidth]{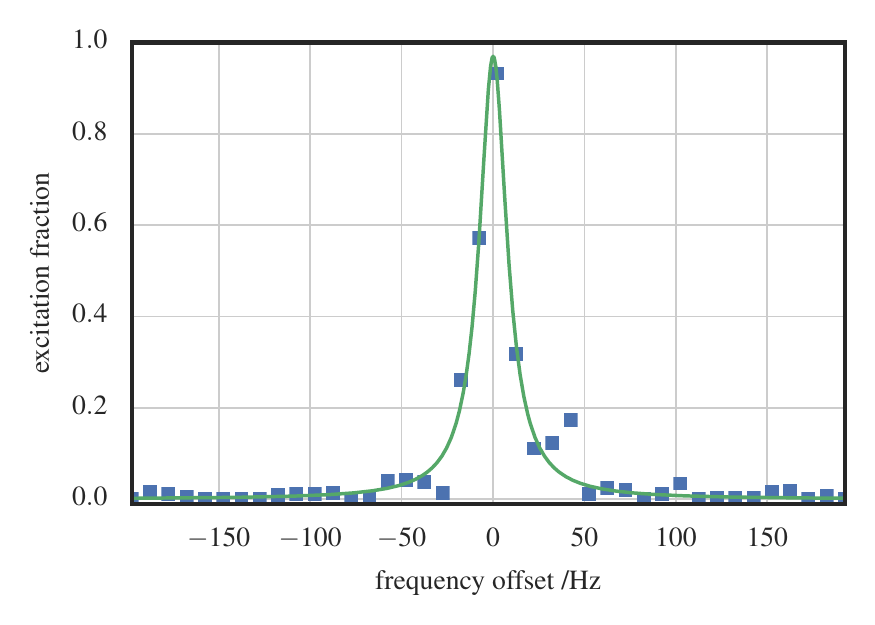}
\caption{Spectroscopy of a single magnetic component of the \ch{^1S_0} -- \ch{^3P_0} transition with spin-polarized atoms.
Blue points are experimental data from a single scan of the resonance. The green line is a lorentzian fit with linewidth \SI{18}{Hz}.
}
\label{fig:spectroscopy}
\end{figure}

The clock transition is observed with a typical linewidth of \SI{20}{Hz} (see for example \cref{fig:spectroscopy}).
The linewidth is limited by excessive clock-laser frequency noise as the seismic noise on the ultrastable cavity is not properly cancelled by the vibration-insensitive supports and by the passive isolation platform.

\begin{figure}
\includegraphics[width=\figwidth]{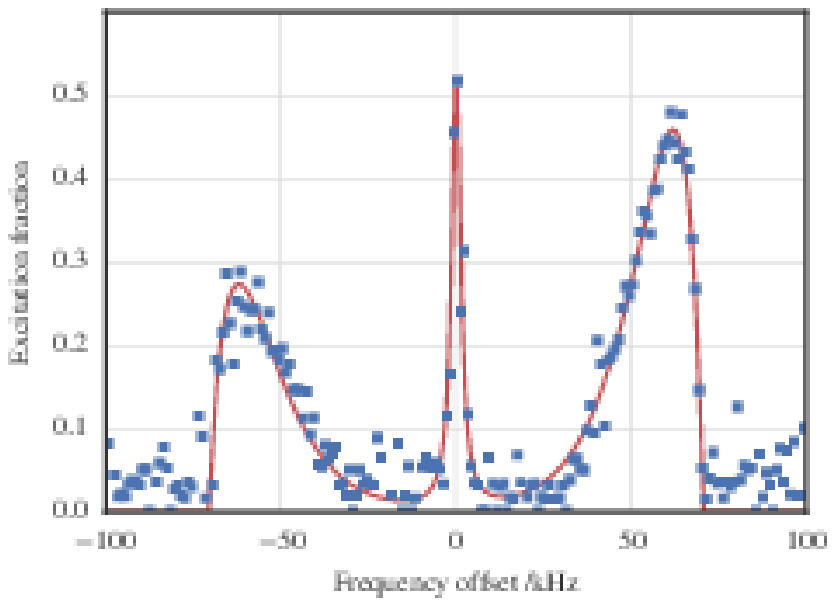}
\caption{Single scan spectroscopy of the clock transition in the optical lattice using \num{1e4} atoms and a Rabi pulse of \SI{80}{ms}.
Blue dots are experimental data. 
The red line is a fit with the combined shape expected for the carrier and the sidebands \cite{Blatt2009}.
From the fit we deduce a longitudinal trap frequency of \SI{68.6(3)}{kHz} corresponding to a trap depth of $\num{284(3)}E\ped{r}$.
From the ratio between the red and blue sideband we can extract the longitudinal atomic temperature in the trap of \SI{7.0(14)}{\micro K}.
}
\label{fig:sidebands}
\end{figure}

To gain a better understanding of the optical trapping conditions we performed broadband spectroscopy and observed the sideband spectrum (\cref{fig:sidebands}). 
From the spectroscopy results and by fitting the spectrum we extracted the trap depth $U_0$ as the longitudinal motional frequency $\nu_z$ is given by as: 
$h \nu_z = 2 E_\mathrm{R} \sqrt{U_0 / E_\mathrm{R}} $
Furthermore the longitudinal temperature of the atoms $T_z$ can be measured 
from the ratio $b\ped{br}$ between the excitation fraction of the blue and red sideband $k\ped{B} T_z={h\nu_z}/{\ln(b\ped{br})}$ \cite{Blatt2009}, where $k_B$ is the Boltzmann constant. 
For example \cref{fig:sidebands} shows a trap depth of $\num{284(3)}E\ped{r}$ and a temperature of \SI{7.0(14)}{\micro K}.
Typically we operate the optical trap at a depth of 200 $E\ped{r}$ at a temperature of \SI{7(3)}{\micro K}.

\subsection {Clock operation}

The clock laser is locked to the average frequency of the two $m_F \!=\! \pm 1/2$ components removing the first-order Zeeman shift  as well as the vector light shift. 
We alternatively interrogate the side of each transition to derive two error signals and we use an independent lock for each. 
In total four interrogations are performed: two for the $m_F \!=\! +1/2$ state and two for the $m_F \!=\! -1/2$ state set at the transitions sides.
Addressing of the four points is controlled by frequency shifting the clock AOM (\cref{fig:setup}), controlling also the Rabi spectroscopy by pulsing the clock laser frequency. 
A computer program sets and records the frequencies of the two transitions.
We then post-process the data to extract the average and splitting of the two resonances.

\begin{figure}
\includegraphics[width=\figwidth]{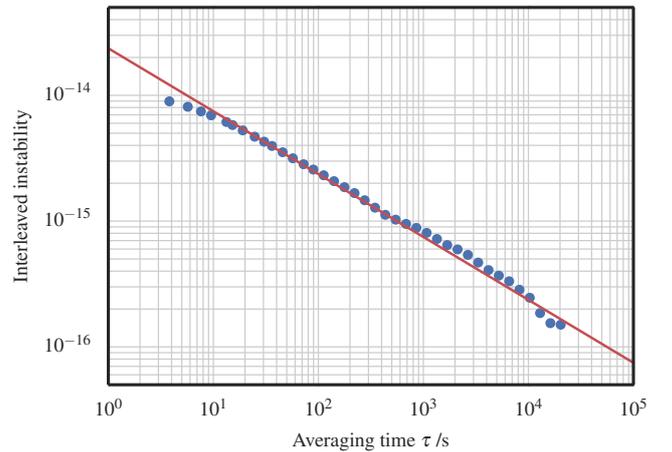}
\caption{Example of the instability of the interleaved measurement used to evaluate the induced lattice light shift. Blue dots are the overlapping Allan deviation of the frequency difference between the lattice at high power and at low power.
The red line corresponds to the white frequency noise as $\num{2.3e-14} /\sqrt{\tau/\si{s}}$.
The total measurement time was \SI{22.5}{h}.
}
\label{fig:interleaved}
\end{figure}

To evaluate systematic shifts we interleave two clock sequences for a total of four independent locks.
We then post-process the data to measure the frequency difference  between the two interleaved states interpolating at common timetags.
\Cref{fig:interleaved}  shows an example of the instability of an interleaved measurement for a total measurement time of  \SI{22.5}{h}.
The instability is compatible with white frequency noise as $\num{2.3e-14} /\sqrt{\tau/\si{s}}$ and is limited by the same cavity noise affecting the linewidth.

\section{Systematic frequency shifts}

\Cref{tab:yb} shows the uncertainty budget for the \ch{^{171}Yb} optical lattice frequency standard.
The fractional uncertainty for the measurement reported here is \num{1.5e-16}.
The contributions of the systematic effects will be discussed below.

\begin{table*}
\caption{Uncertainty budget for the Yb optical lattice clock.}
\setlength{\tabcolsep}{9pt}
\label{tab:yb}
\centering
\begin{tabular}{lS[table-format = 2.4]S[table-format = 1.4]@{\hskip 18pt}S[table-format = 4.2]S[table-format = 2.1]}
\toprule
Effect 	&{Shift.\ /Hz}	&{Unc.\ /Hz}		&{Rel.\ Shift$\times$\num{e17}}	&{Rel.\ Unc.$\times$\num{e17}}\\
\midrule
{Lattice polarizability} 	&-0.02	&0.04	&-4	&8\\
{Nonlinear lattice shifts} 	&-0.07	&0.05	&-12	&10\\
{Density shift} 	&-0.01	&0.03	&-2	&6\\
{Zeeman shift}	&-0.14	&0.02	&-27	&4\\
{BBR room} 		&-1.219	&0.013	&-235.3	&2.5\\
{BBR oven} 		&-0.011	&0.004	&-2.1	&0.8\\

{Probe light shift} 	&0.005	&0.018	&1.	&3.5\\
{Doppler shift}		&{--}	&0.03	&{--}	&5\\
{Background gas shift} 	&{--}	&0.008	&{--}	&1.5\\
{Static Stark shift} 	&{--}	&0.005	&{--}	&1\\
{Servo error} 	&{--}	&0.006	&{--}	&1\\
{Tunnelling} 	&{--}	&0.002	&{--}	&0.4\\
{Line pulling} 	&{--}	&0.002	&{--}	&0.4\\
{AOM switching} 	&{--}	&0.002	&{--}	&0.4\\
{Fiber links} 	&{--}	&0.0005	&{--}	&0.1\\
{Gravitational redshift} &13.474	&0.003	&2601.2	&0.6\\
\midrule
\textbf{Total}	&12.01	&0.08	&2320	&16\\
\bottomrule
\end{tabular}
\end{table*}

\subsection{Lattice light shifts}

The induced lattice light shift can be written as \cite{Katori2015,Nemitz2016}:
\begin{multline}\label{eq:lattice}
\Delta\nu\ped{ls} = 
- \left( a \Delta\nu +\frac{3}{4} d (2n^2 + 2n + 1 )\right) \frac{U\ped{e}}{E\ped{r}}  
 - d {\left(\frac{U\ped{e}}{E\ped{r}}\right)}^2 +\\
+ (a \Delta\nu - b)\left(n+\frac{1}{2}\right) {\left(\frac{U\ped{e}}{E\ped{r}}\right)}^{1/2}   
+ d (2 n + 1) {\left(\frac{U\ped{e}}{E\ped{r}}\right)}^{3/2} 
\end{multline}
where $\Delta\nu$ is the lattice frequency shift from the electric-dipole (E1) magic wavelength $\nu_{E1}$ and $U\ped{e} = \xi U_0 $ is the effective trap depth that takes into account the reduced laser intensity seen by the atoms. The coefficient $a$ is the linear slope of the polarizability per lattice frequency, while $b$ is a multipolar polarizability, and $d$ the hyperpolarizability coefficient. $n$ is the average motional state occupation in the lattice.

We determine the linear part of the shift and the magic wavelength by interleaving two clock cycles operating at different lattice intensities.
Experimentally we measure the frequency difference between $U_0 = U\ped{L}$ and $U_0 = U\ped{H}$ and calculate the resulting shift at $U_0 = U\ped{L}$ only assuming a linear shift.
We then correct these values and calculate the hyperpolarizability and multipolar shifts using \cref{eq:lattice} and  the values $b=\SI{-0.68(71)}{mHz}$ and $d=\SI{-1.9(8)}{\micro Hz}$ known for Yb \cite{Nemitz2016, Barber2006}.

The lattice intensity is modulated changing the diffraction efficiency of the AOM before the fiber delivering the laser to the atoms.
This method preserves the frequency spectrum of the titanium-sapphire laser between high and low power.
Atoms are always loaded at low power with $U\ped{L} \approx 200 E\ped{r}$ while to measure at high power the trap depth is abruptly raised before spectroscopy to $U\ped{H} \approx 300 E\ped{r}$ in about \SI{0.1}{ms}.

We use a fast photodiode (bandwidth \SI{3}{GHz}) to check that the titanium-sapphire laser is in single mode.
The broadband emission spectrum of the laser is not filtered out for the present measurement.


We measured the shift at the working point  $\nu\ped{l} =\SI{394798.238}{GHz}$ between $U\ped{L} = \num{196(4)}E\ped{r}$ and $U\ped{H} = \num{348(5)}E\ped{r}$.
The effective shift at $U\ped{L}$ is  \SI{-0.02(4)}{Hz}. 


The lattice depths, $\xi$, and $n$ were determined by sidebands spectroscopy (\cref{fig:sidebands}).
For our trap configuration, $\xi=\num{0.8(1)}$ and $n=\num{2.1(10)}$ at $U\ped{L}$ and  $n=\num{1.5(8)}$ at  $U\ped{H}$.
We correct the nonlinearity between $U\ped{H}$ and $U\ped{L}$ in the effective shift and  calculate the hyperpolarizability and multipolar shifts using \cref{eq:lattice}. 
These two contributions are  \SI{-0.13(6)}{Hz} and \SI{0.064(32)}{Hz} and sum to \SI{-0.07(5)}{Hz}.
This is the largest contribution to the Yb uncertainty and is due to the uncertainty in the multipolar and hyperpolarizability coefficients at high trap depths.

\begin{figure}
\includegraphics[width=\figwidth]{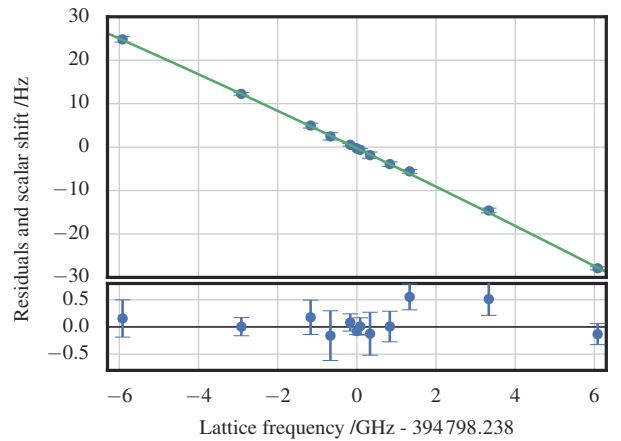}
\caption{Scalar lattice shift at $U_0 = 200 E\ped{r}$ measured by interleaving between a trap depth of $\num{208(1)}E\ped{r}$ and $\num{333(2)}E\ped{r}$ as a function of lattice frequency and applying a small correction for multipolar and hyperpolarizability effects.
Blue points show experimental data while the green line is a quadratic fit.
Bottom panel shows the residuals of the fit. }
\label{fig:lattice}
\end{figure}

We also measured the lattice shift as a function of the lattice frequency.
\Cref{fig:lattice} shows the scalar shift measured between a trap depth of $U\ped{L} = \num{208(2)}E\ped{r}$ and $U\ped{H} = \num{333(4)}E\ped{r}$ as a function of lattice frequency.
Fitting the data with a second order Taylor expansion around the magic wavelength the linear coefficient is $a = \SI{27(3)}{mHz/GHz}$ and the quadratic coefficient is $c = \SI{0.16(3)}{mHz/GHz^2}$ to give a shift $-c {\Delta\nu}^2 U\ped{e}/E\ped{r}$.
Uncertainty is limited by $\xi$ and not by the statistical uncertainty in the fit.

We calculated the $E1$-magic frequency from the measured linear coefficient $a$, the measured shift at the working point, and the nonlinear corrections.
For our system it is \SI{394798.205(17)}{GHz}. This measurement of the magic frequency and the ones reported before \cite{Lemke2009, Nemitz2016} are not consistent with each other as they could depend on the spectral features of the titanium-sapphire lasers if not filtered \cite{Targat2012}.
Our data is consistent with a constant shift coming from the spectrum of the lasers.





\subsection{Density shift}
In \ch{^{171}Yb} optical frequency standards the density-dependent shift is dominated by $p$-wave cold collision \cite{Lemke2011} while $s$-wave scattering can arise from inhomogeneity of the Rabi excitation \cite{Campbell2009}.
To evaluate the density shift we run interleaved measurements varying the number of atoms.
This is modulated by changing the duration of the slower beam pulse at \SI{399}{nm} during the first stage MOT.
This method impacts on the atomic density without changing the trapping conditions and we assume the atomic density to be proportional to the atoms number.
The trap region has a radius of \SI{45}{\micro m}, given by the lattice waist, and is \SI{0.5}{mm} long, given by the size of the second stage MOT.
For our setup an atomic density of $\rho_0 = \SI{4e14}{m^{-3}}$ corresponds to about 1 atom per lattice well.
The absolute atom number has a fractional uncertainty of \SI{20}{\percent} but the relative density measurements have an uncertainty of about \SI{1}{\percent}.
During our measurements, we varied the atomic density from $\rho\simeq0.1\rho_0$ to $\rho\simeq3\rho_0$.

\begin{figure}
\includegraphics[width=3.2in]{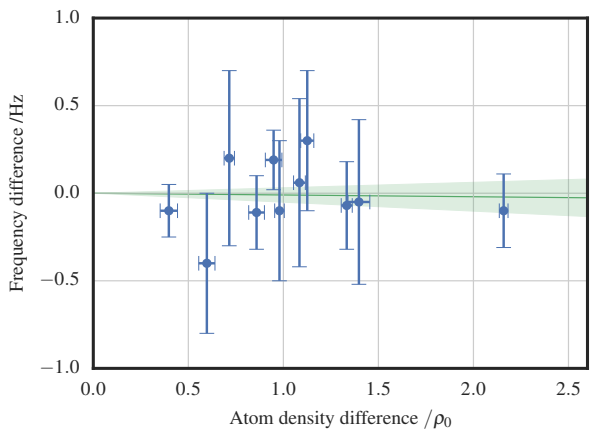}
\caption{Density shift measured interleaving between high and low atom numbers with an excitation fraction of \num{0.39(2)}.
Blue points mark the experimental data with their uncertainty.
The green line and the shaded region show a linear fit with its statistical uncertainty.
 }
\label{fig:density}
\end{figure}

The density shift depends on the details of the Rabi pulse used for spectroscopy and whether the atomic sample is spin-polarized \cite{Ludlow2011}.
\Cref{fig:density} shows the density shift measurement as a function of the atomic density difference interleaving between high and low number of atoms with a spin-polarized sample and a final excitation fraction (measured after the Rabi pulse) of \num{0.39(2)}, given by our choice of the locking point on the resonance. 
We fit the data with a linear function with no offset and measured a slope of $\SI{-0.01(4)}{Hz}/\rho_0$.
At a different locking point and an final excitation fraction of \num{0.34(2)} we measured in a similar way a slope of $\SI{-0.4(2)}{Hz}/\rho_0$.
The dependence on excitation fraction is predicted by  $p$-wave collisions \cite{Ludlow2011}.
For our conditions $\rho\simeq0.7\rho_0$ and excitation fraction \num{0.39(2)} the density shift is \SI{-0.01(3)}{Hz}.

\subsection{Zeeman shift}
To measure the clock transition we apply a magnetic field and the Zeeman effect split the two $\pi$-transition resonances between $m_F \!=\! \pm1/2$ states.
This split gives a continuous measurement of the magnetic field experienced by the atoms while running the experiment.
The operational Zeeman shift for the two components from the center is \SI{290.1(3)}{Hz} corresponding to $B=\SI{138(1)}{\micro T}$  \cite{Lemke2009}.
We observed no drift of the magnetic field down to \SI{0.1}{\micro T/h}.
Taking the average frequency of the two $m_F \!=\! \pm1/2$ magnetic components removes the first-order Zeeman shift.
We calculated a quadratic Zeeman shift of \SI{-0.14(2)}{Hz} using the quadratic Zeeman coefficient known for Yb \SI{7(1)}{Hz/mT} \cite{Lemke2009}.

\subsection{Black-body radiation shift}

The black-body radiation (BBR) emitted by bodies at temperature $T \neq \SI{0}{K}$ surrounding the atomic cloud induces a systematic Stark shift of the clock transition frequency according to \cite{Sherman2012}

\begin{equation}
	\Delta \nu_\text{BBR} = -\frac{1}{2} (\alpha^{(0)}_{\ch{^3P_0}}-\alpha^{(0)}_{\ch{^1S_0}}) \braket{E^2}_T ( 1 + \eta_\text{clock}(T)).
\end{equation} 

The coefficients $\alpha^{(0)}_{\ket{i}}$ are the static polarizabilities of the clock transition ground and excited states, $\braket{E^2}_T$ is the average electric field radiated by a black body at temperature $T$, $\eta_\text{clock}$ is a dynamic correction. For \ch{^{171}Yb}  is $\Delta\alpha^{(0)} = (\alpha^{(0)}_{\ch{^3P_0}}-\alpha^{(0)}_{\ch{^1S_0}}) = \SI{36.2612(7)}{kHz(kV cm)^{-2}}$ and $\eta(T=\SI{300}{K}) = \num{0.0179(5)}$ \cite{Sherman2012, Beloy2014}.

We kept into account the vacuum package and the Yb oven to evaluate the BBR shift. 

\begin{figure}
\includegraphics[width=3.2in]{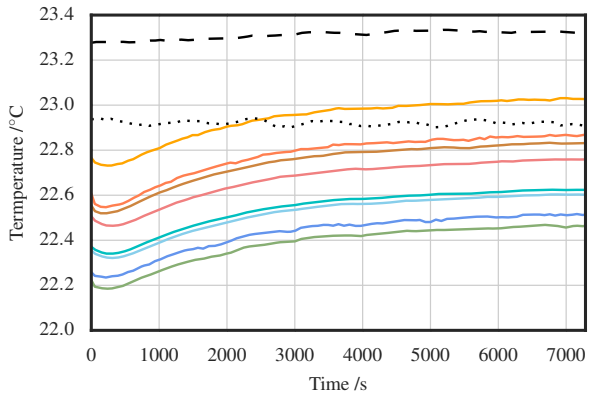}
\caption{Example of the temperatures of the ten Pt1000 platinum resistance thermometers surrounding the vacuum chamber recorded during a single measurement run. Solid colored lines mark the temperature of the eight resistance thermometers on the trapping chamber. The black dashed line is the temperature recorded close to the atomic oven while the black dotted line is the temperature measured far away.}
\label{fig:temperatures}
\end{figure}

The vacuum package temperature has been continuously monitored at ten different positions by platinum resistance thermometers Pt1000.
Gradients are caused by the power dissipated by the MOT coils, by the temperature of the water cooling, and by the temperature of the oven.
Eight Pt1000 are placed on the chamber enclosing the atom trapping region while other two are placed respectively close to the oven and far away from it.
The Pt1000 calibration uncertainty is \SI{0.2}{K} while the 4-wires resistance reading uncertainty is \SI{0.06}{K}. 
The temperature spread between different thermometers lays typically within \SI{2}{K}.
We observed bigger inhomegeneities during transients at the start of the experiment, while changing the magnetic field gradient in the MOT coils, or  increasing the oven temperature.
\Cref{fig:temperatures} shows an example of the recorded temperatures during a measurement.


We assumed a rectangular probability density function for the temperature seen by the atoms, taking as average value $T = (T\ped{max} + T\ped{min})/2$ and as uncertainty $(T\ped{max} - T\ped{min})/\sqrt{12}$ where $T\ped{max}$ and $T\ped{min}$ are the maximum and minimum temperature recorded during a single run \cite{BIPM2008a}.
The uncertainty coming from the temperature inhomogeneity for this measurement is \SI{0.8}{K}.
This approach leads to a  BBR shift due to the environment temperature of \SI{-1.219(13)}{Hz}.

The BBR emitted by the oven reaches the atom trapping region by passing through a copper vacuum-differential tube which is \SI{15}{cm} long and has a \SI{0.5}{cm} inner diameter. Thus the atoms are affected by both the radiation from the oven with direct line of sight and the radiation reaching the atoms after being reflected on the tube inner surface. A ray tracing analysis has been performed to calculate the effective solid angle under which the atomic cloud sees direct and reflected radiations. In the effective solid angle calculation the emissivity of the tube material is also kept into account as $\epsilon\ped{Cu} = \num{0.4(2)}$ and two reflections are relevant in the shift calculation.

The oven temperature of \SI{400(10)}{\celsius} has been continuously monitored using a thermocouple. 
The uncertainty considers the thermocouple accuracy and differences in temperature between the oven tip and the sensor position.
The BBR shift contribution from the oven is then calculated to be \SI{-0.011(4)}{Hz}.


For the first measurements in this campaign we had also a window heated at \SI{230(5)}{\celsius} with direct sight to the atoms.
For the effective solid angle underlying the hot window, we modeled up to 5 reflections from stainless steel surfaces (fittings to the window) and 1 reflection from a polished aluminum surface (main chamber), with the material emissivity $\epsilon\ped{steel} = \num{0.5(3)}$ and $\epsilon\ped{Al} = \num{0.13(10)}$.
The resulting shift was \SI{-0.12(15)}{Hz}. 
After February 2016 we replaced  the hot window with a cold intra-vacuum mirror thus removing this shift and source of uncertainty.
The contribution of the hot window BBR to the final relative uncertainty for the measurement presented later is at the  \num{5e-17} level as we averaged data with and without this shift.

\subsection{Probe light shift}
The interaction of the probe laser with the atoms during Rabi spectroscopy causes a Stark shift on the transition.
We have calculated this shift by using the published value \cite{Lemke2009} and noticing that for a $\pi$ pulse the laser intensity scales as the inverse of the square of the Rabi time.
From this rescaling it follows that the Stark shift induced by the probe laser is $\SI{0.005(18)}{Hz}$.

\subsection{Gravitational redshift}
The difference between the height of the Yb atoms and the integrated height of the Cs atoms along the free-flight trajectory is $h(\text{Yb-Cs}) = \SI{0.15(2)}{m}$.
The gravitational redshift for the comparison is then \num{1.6(2)e-17}.

The gravitational potential relative to the geoid at INRIM's laboratories  was measured in 2013  in the context of the European project ``International timescales with optical clocks'' (ITOC) \cite{Margolis2013, Denker2015a}.
The measurement was performed both by a levelling approach and by a Global Navigation Satellite System/geoid (GNSS/geoid) approach \cite{Calonico2007}.
The gravitational potential at the Yb atoms position is $C(\text{Yb})\ped{lev} = \SI{2336.25(35)}{m^2/s^2}$ as determined by levelling 
and  $C(\text{Yb})\ped{GNSS/geoid} = \SI{2336.48(27)}{m^2/s^2}$ as determined by GNSS/geoid \cite{Denker2015b} relative
to the International Earth Rotation and Reference Systems Service (IERS) 2010 conventional reference potential $W_0 = \SI{62636856.0(5)}{m^2/s^2}$ \cite{Petit2010, Groten2004}.
Geometric levelling is accurate over short distances but is susceptible to systematic errors which may accumulate over distances $>\SI{1000}{km}$ \cite{Denker2013}.
Therefore the GNSS/geoid approach is beneficial in the comparison between different laboratories by satellite or long-haul optical fiber links.
In this case the GNSS/geoid value gives a gravitational redshift on the Yb atoms of \SI{13.474(3)}{Hz} with a fractional uncertainty of \num{6e-18} limited by the uncertainty  on the reference $W_0$.

The corresponding numbers for ITCsF2 are $C(\text{Cs})\ped{lev} = \SI{2334.8(10)}{m^2/s^2}$ and $C(\text{Cs})\ped{GNSS/geoid} = \SI{2335.0(10)}{m^2/s^2}$.
Using the GNSS/geoid value the redshift for ITCsF2 is then \num{2.599(1)e-14}.


\subsection{Other shifts}

We measured that the phase-stabilized fiber links between the cavity and the atoms contribute a fractional uncertainty of \num{1e-18} after \SI{1000}{s} of integration time.
However the fiber link to the atoms is referenced to a mirror placed \SI{30}{cm} from the lattice back-reflector, that is the reference for the lattice.
We discovered some phase-noise between the two references which include noise syncronous with the clock cycle (e.g., MOT coils switching).
We measured a fractional contribution of \num{5e-17} that we take as uncertainty coming from the Doppler shift.

A static electric field would cause a Stark shift on the clock transition \cite{Sherman2012}.
Our vacuum chamber is made of aluminum and is a Faraday cage from external electric fields. However it is possible for electric charges to be present on the windows of the vacuum chamber \cite{Lodewyck2012}. In our setup two fused silica windows are \SI{25}{mm} away from the atoms and have a radius of \SI{38}{mm} and a thickness of \SI{12}{mm}.
We estimated a capacitance of \SI{0.6}{pF} between the two surfaces and a time constant of $\sim\SI{50}{d}$ assuming a high resistivity of \SI{e18}{\ohm m} for fused silica  \cite{Lodewyck2012}.
We have operated our vacuum system for more than 3 years and possible charges present at the beginning should have decayed to give a Stark shift $< \num{1e-17}$ that we take as uncertainty.

The frequency shift  due to collisions with the hot atomic beam and background gas can be related to the lifetime of the atoms in the trap \cite{Gibble2013}. From our observed lifetime of \SI{2.7}{s} and the $C_6$ coefficients for Yb dimers \cite{Porsev2014} we assign a shift of $\SI{8}{mHz}$ that we take as uncertainty.

Tunnelling of atoms from one lattice site to the other leads to a broadening of the band structure of the lattice \cite{Lemonde2005} and in our trap is suppressed by the deep potential.
Given our trap depth of $U_0 = 200 E\ped{r}$ and temperature of $T = \SI{7(3)}{\micro K}$ we calculated that \SI{70}{\%} of the atoms occupy the lowest three band in the lattice \cite{Lemonde2005}. We then take the width of the third band \SI{2}{mHz} as uncertainty .

We calculated the line pulling from the lattice sidebands to be \SI{2}{mHz} that we take as uncertainty.
The line pulling from the $\pi$ and $\sigma$ transitions between $m_F \!=\! \pm1/2$ states is negligible because of spin-polarization of the atoms and the linear polarization of the probe light.

We measured a possible error in the servo loops taking the average of the error signal measured during the campaign that gives a fractional uncertainty of \num{1e-17}.
We switch the frequency of the AOM keeping a constant power to avoid thermal effects.
We did not detect any shift up to \SI{2}{mHz} by heterodyne interferometry, that we take as uncertainty.
The switching of the AOM to apply the Rabi pulse introduces a phase jump in the phase-stabilization.
We calculated a frequency shift of \SI{0.5}{mHz} \cite{Falke2012} that we take as an uncertainty.

\section{Absolute frequency measurement}
We measured the absolute frequency of the \ch{^{171}Yb} frequency standard relative to the caesium fountain ITCsF2 from January 2016 to the end of May 2016.
The comparison has been performed with a fiber frequency comb with a repetition rate of \SI{250}{MHz} referenced to a hydrogen maser.
The fountain ITCsF2 continuously measured the maser frequency for the whole duration of the campaign.

The light at \SI{578}{nm} resonant with the cavity was sent to the frequency comb through an optical fiber link with active phase-stabilization.
The fiber link to the comb contributes to an uncertainty of \num{1e-18} above \SI{1000}{s} of integration time.
Two different computers measured the frequency of the cavity-stabilized light relative to the frequency comb and the frequency difference between the cavity and the ytterbium atoms of the optical standard.
The frequency comb and the optical standard recorded data at different rates and we averaged and interpolated the data to common timestamps.
Both computers were synchronized to INRIM's Network Time Protocol (NTP) server within \SI{3}{ms}.
Since the maximum observed frequency drift of the cavity was \SI{0.1}{Hz/s} we assume an uncertainty of \num{6e-19} coming from synchronization.

The combs contribution to uncertainty was assessed by measuring the same laser frequency with two independent combs and is \num{1e-16}, limited by the comb electronics noise at an averaging time of \SI{1e4}{s}.
The frequency of the beat note between the cavity-stabilized laser and the comb was measured redundantly to detect cycles slips in the counter.
We discarded all points where the counter readings did not agree at better than \SI{0.2}{Hz}.

The cryogenic fountain ITCsF2 ran in two different atomic density regimes \cite{Levi2014} during the campaign. In the low-density regime ITCsF2 contributes to an uncertainty of \num{3.0e-16} with a stability of $\num{3.6e-13}/\sqrt{\tau/\si{s}}$. In the high-density regime it contributes to an uncertainty of \num{4.8e-16} with a stability of $\num{2.5e-13}/\sqrt{\tau/\si{s}}$. The measurement of the hydrogen maser frequency relative to ITCsF2 was considered only in the  windows of operation of the ytterbium standard.

\begin{figure}
\includegraphics[width=3.2in]{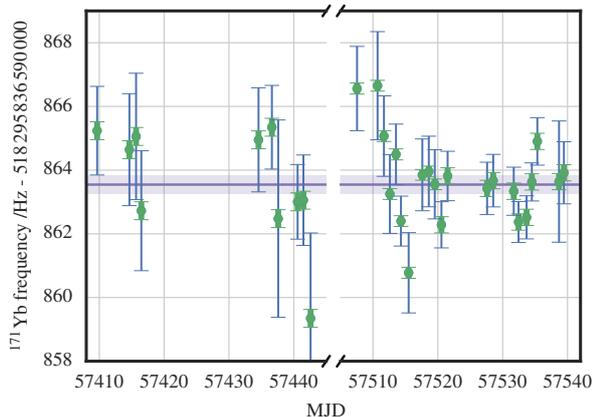}
\caption{Data of the measurement campaign run from January 2016 to May 2016 of the absolute frequency of \ch{^{171}Yb} relative to ITCsF2.
The shaded horizontal region denotes the $1\sigma$ uncertainty for the campaign. 
Blue thin bars mark the combined statistical and systematic uncertainty for each measurement.
Green thick bars mark the combined systematic uncertainty of ITCsF2 and the Yb frequency standard.
}
\label{fig:janmay}
\end{figure}

We took a total of 31 different measurements for a total measurement time of the optical frequency standard of \SI{227}{h}.
\Cref{fig:janmay} shows the different measurements as a function of the Modified Julian Date (MJD).
For each measurement the statistical uncertainty  is dominated by the fountain instability.
The systematic uncertainty varies for each measurement because of the different density values of both the fountain and optical frequency standard and because of the reduced BBR uncertainty contribution on the ytterbium after removing the hot window after the first 10 measurements.
We applied a statistical analysis of the data based on the Gauss-Markov theorem \cite{Luenberger1998, Cox2006} that considers the correlations between the different measurements coming from the systematic shifts \cite{Margolis2015}. The measurement result is $f_{\ch{^{171}Yb}} = \SI{518295836590863.59(31)}{Hz}$ where the uncertainty is detailed in \cref{tab:budget}.
The reduced chi squared with 30 degrees of freedom for the average is $\chi\ped{red}^2 = 1.36$ and is calculated from the complete correlation matrix of the measurements \cite{Margolis2015}. This  chi squared has a $p$-value of \SI{10}{\percent}.
The total fractional uncertainty is \num{5.9e-16} and it is dominated by the uncertainty of the caesium fountain.
The fountain systematic uncertainty contributes for a total of \num{4.0e-16}.
The statistical uncertainty resulted \num{3.4e-16} but we increased this value to \num{3.9e-16}  to account for $\chi\ped{red}^2 > 1$.
A contribution of \num{1.6e-16} comes from the ytterbium optical frequency standard as explained above.

\begin{table}
\caption{Uncertainty budget for the Yb-ITCsF2 comparison}
\label{tab:budget}
\centering
\begin{tabular}{lSS}
\toprule
Contribution 	&{Unc.\! /Hz}		&{Frac.\!\! Unc.$\times$\num{e16}}\\
\midrule
{ITCsF2}		&0.21	&4.0\\
{Statistics}	&0.20	&3.9\\
{Yb}		&0.08	&1.6\\
{Comb}		&0.05	&1.\\
{Grav. redshift}	&0.001	&0.02\\
{Fiber link}	&0.0005	&0.01\\
{Synchronization} 	&0.0003	&0.006\\
\midrule
\textbf{Total} 	&0.31	&5.9\\
\bottomrule
\end{tabular}
\end{table}

\begin{figure}
\includegraphics[width=3.2in]{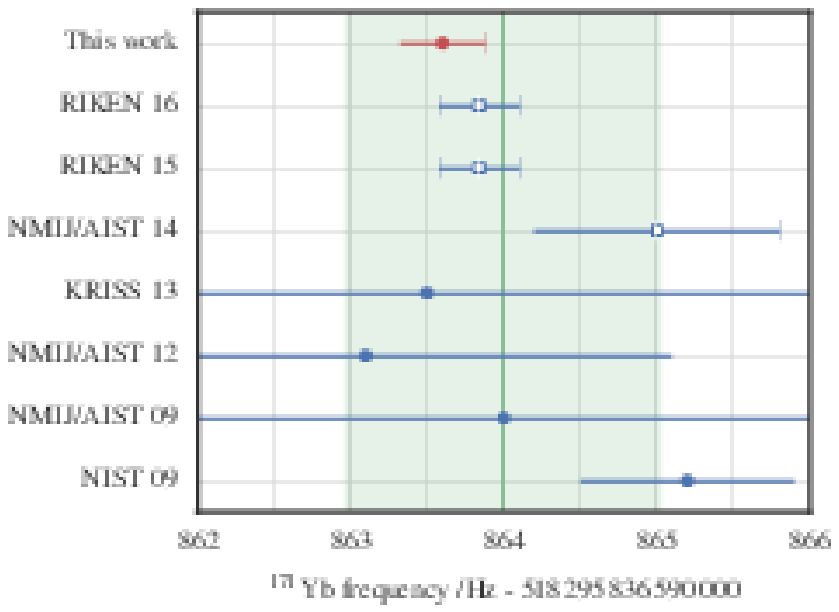}
\caption{Absolute frequency measurement of the \ch{^1S_0} -- \ch{^3P_0} transition of \ch{^{171}Yb} measured by different laboratories and years: NIST \cite{Lemke2009}, NMIJ \cite{Kohno2009, Yasuda2012, Akamatsu2014}, KRISS \cite{Park2013}, RIKEN \cite{Takamoto2015, Nemitz2016}. The shaded region gives the recommended frequency for \ch{^{171}Yb} as a secondary representation of the second of CIPM 2015, with its uncertainty. 
Open squares denote absolute frequencies deduced from \ch{^{171}Yb}/\ch{^{87}Sr} ratio measurements and the recommended value of \ch{^{87}Sr} as a secondary representation of the second with its uncertainty. 
Circles denote measures relative to the realization of the second by caesium standards.
}
\label{fig:comparison}
\end{figure}

\Cref{fig:comparison} compares our value with previous measurements and the recommended frequency for \ch{^{171}Yb} as a secondary representation of the second  $f_{\ch{^{171}Yb}}(\text{CIPM2015}) = \SI{518295836590864.0(10)}{Hz}$ \cite{CIPM2015}.
Previous measurements include values measured relative to the realization of the second by caesium standards \cite{Lemke2009, Kohno2009, Yasuda2012, Park2013} and absolute frequencies deduced from optical ratio measurements with \ch{^{87}Sr} frequency standards \cite{Akamatsu2014,Takamoto2015, Nemitz2016}.
To deduce these values we used the recommended frequency of \ch{^{87}Sr}  $f_{\ch{^{87}Sr}}(\text{CIPM2015}) = \SI{429228004229873.2(2)}{Hz}$ as a secondary representation of the second with its uncertainty of \num{5e-16} \cite{CIPM2015}.

\section{Conclusions}
We report the absolute frequency measurement of the unperturbed \ch{^1S_0} -- \ch{^3P_0} transition of \ch{^{171}Yb} with a fractional uncertainty of \num{5.9e-16} performed in an optical lattice frequency standard and limited by the caesium fountain. The meausurement result is $f_{\ch{^{171}Yb}} = \SI{518295836590863.59(31)}{Hz}$ where the number in parentheses is the numerical value of the standard uncertainty referred to the corresponding last digit of the quoted result.
Our measurement is independent from previous measurements of the transition and the result is in agreement with the recommended value for \ch{^{171}Yb} as a secondary representation of the SI second.
This measurement of the ytterbium transition is the first in Europe after previous measurements achieved in North America and Asia and is important in view of a possible redefinition of the SI second based on optical frequency standards.

We characterized the ytterbium optical frequency standard and evaluated its fractional uncertainty for the presented measurement to \num{1.6e-16}.
We expect that this uncertainty can be reduced to the \num{1e-17} level in the near future.
In this measurement the Yb standard is limited by the poor short-term stability due to excessive noise on the ultrastable cavity.
Improving the cavity environment and the performance of the clock laser will allow us to reduce the uncertainty measured by interleaving to \num{1e-17} (lattice shift and density shift).
The nonlinear part of the lattice shift can be reduced by working with smaller lattice depths and by improving the measurement of multipolar and hyperpolarizability effects.
The Zeeman shift can be reduced by working with smaller magnetic fields.
The Doppler shift can be easily cancelled by referencing the clock laser to the lattice mirror.
The implementation of a mechanical shutter in the atomic beam path will further reduce the effects of the background collisions and the BBR shift coming from the hot atomic oven.
The static Stark shift can be better evaluated by placing electrodes on the vacuum chamber windows.
The final limitation of the current setup is the temperature uniformity of the vacuum chamber leading to a total BBR contribution with a fractional uncertainty of \num{2.5e-17}.
Improving the thermal isolation from  environment and heat sources (e.g., the MOT coils) and waiting for thermal transient to decay will be beneficial in further reducing this contribution to \num{1e-17}.

\section*{Acknowledgments}
The authors would like to acknowledge Aldo Godone, Elio Bertacco, Claudio Calosso and Massimo Zucco for help and useful discussions.
The authors acknowledge funding
from the European Metrology Research Program (EMRP) Project SIB55-ITOC, 
from the Regione Piemonte YTRO Project,
from the Space Optical Clocks project of the European Union Seventh Framework Programme (FP7/2007-2013 grant agreement 263500),
from the Italian Ministry of Education, University and Research Project PRIN2012 AQUASIM,
from the Innovative Training Network (ITN) Future Atomic Clock Technology (FACT),
and from the European Metrology Program for Innovation and Research (EMPIR) project 15SIB03 OC18.  
The EMRP is jointly funded by the EMRP participating countries within EURAMET and the European Union.
This project has received funding from the EMPIR programm co-financed by the Participating States and from the European Union's Horizon 2020 research and innovation programme.

\bibliographystyle{iopart-num} 
\bibliography{/home/marco/Documents/main}

\providecommand{\newblock}{}
\begin{thebibliography}{10}
\expandafter\ifx\csname url\endcsname\relax
  \def\url#1{{\tt #1}}\fi
\expandafter\ifx\csname urlprefix\endcsname\relax\def\urlprefix{URL }\fi
\providecommand{\eprint}[2][]{\url{#2}}

\bibitem{Levi2014}
Levi F, Calonico D, Calosso C~E, Godone A, Micalizio S and Costanzo G~A 2014
  {\em Metrologia\/} {\bf 51} 270
  \urlprefix\url{http://stacks.iop.org/0026-1394/51/i=3/a=270}

\bibitem{Heavner2014}
Heavner T~P, Donley E~A, Levi F, Costanzo G, Parker T~E, Shirley J~H, Ashby N,
  Barlow S and Jefferts S~R 2014 {\em Metrologia\/} {\bf 51} 174
  \urlprefix\url{http://stacks.iop.org/0026-1394/51/i=3/a=174}

\bibitem{Guena2012}
Guena J, Abgrall M, Rovera D, Laurent P, Chupin B, Lours M, Santarelli G,
  Rosenbusch P, Tobar M~E, Li R, Gibble K, Clairon A and Bize S 2012 {\em IEEE
  Transactions on Ultrasonics, Ferroelectrics, and Frequency Control\/} {\bf
  59} 391--409 ISSN 0885-3010

\bibitem{Ludlow2015}
Ludlow A~D, Boyd M~M, Ye J, Peik E and Schmidt P~O 2015 {\em Rev. Mod. Phys.\/}
  {\bf 87}(2) 637--701
  \urlprefix\url{http://link.aps.org/doi/10.1103/RevModPhys.87.637}

\bibitem{Ushijima2015}
Ushijima I, Takamoto M, Das M, Ohkubo T and Katori H 2015 {\em Nat Photon\/}
  {\bf 9} 185--189 ISSN 1749-4885
  \urlprefix\url{http://dx.doi.org/10.1038/nphoton.2015.5}

\bibitem{Yamanaka2015}
Yamanaka K, Ohmae N, Ushijima I, Takamoto M and Katori H 2015 {\em Phys. Rev.
  Lett.\/} {\bf 114}(23) 230801
  \urlprefix\url{http://link.aps.org/doi/10.1103/PhysRevLett.114.230801}

\bibitem{Targat2013}
Le~Targat R, Lorini L, Le~Coq Y, Zawada M, Gu\'ena J, Abgrall M, Gurov M,
  Rosenbusch P, Rovera D~G, Nag\'orny B, Gartman R, Westergaard P~G, Tobar M~E,
  Lours M, Santarelli G, Clairon A, Bize S, Laurent P, Lemonde P and Lodewyck J
  2013 {\em Nat Commun\/} {\bf 4} --
  \urlprefix\url{http://dx.doi.org/10.1038/ncomms3109}

\bibitem{Lemke2009}
Lemke N~D, Ludlow A~D, Barber Z~W, Fortier T~M, Diddams S~A, Jiang Y, Jefferts
  S~R, Heavner T~P, Parker T~E and Oates C~W 2009 {\em Phys. Rev. Lett.\/} {\bf
  103} 063001

\bibitem{Huntemann2016}
Huntemann N, Sanner C, Lipphardt B, Tamm C and Peik E 2016 {\em Phys. Rev.
  Lett.\/} {\bf 116}(6) 063001
  \urlprefix\url{http://link.aps.org/doi/10.1103/PhysRevLett.116.063001}

\bibitem{Madej2012}
Madej A~A, Dub\'e P, Zhou Z, Bernard J~E and Gertsvolf M 2012 {\em Phys. Rev.
  Lett.\/} {\bf 109}(20) 203002
  \urlprefix\url{http://link.aps.org/doi/10.1103/PhysRevLett.109.203002}

\bibitem{Rosenband2008}
Rosenband T, Hume D~B, Schmidt P~O, Chou C~W, Brusch A, Lorini L, Oskay W~H,
  Drullinger R~E, Fortier T~M, Stalnaker J~E, Diddams S~A, Swann W~C, Newbury
  N~R, Itano W~M, Wineland D~J and Bergquist J~C 2008 {\em Science\/} {\bf 319}
  1808--1812
  \urlprefix\url{http://www.sciencemag.org/content/319/5871/1808.abstract}

\bibitem{Lorini2008}
Lorini L, Ashby N, Brusch A, Diddams S, Drullinger R, Eason E, Fortier T,
  Hastings P, Heavner T, Hume D, Itano W, Jefferts S, Newbury N, Parker T,
  Rosenband T, Stalnaker J, Swann W, Wineland D and Bergquist J 2008 {\em The
  European Physical Journal Special Topics\/} {\bf 163} 19--35 ISSN 1951-6401
  \urlprefix\url{http://dx.doi.org/10.1140/epjst/e2008-00807-7}

\bibitem{Huang2016}
Huang Y, Guan H, Liu P, Bian W, Ma L, Liang K, Li T and Gao K 2016 {\em Phys.
  Rev. Lett.\/} {\bf 116}(1) 013001
  \urlprefix\url{http://link.aps.org/doi/10.1103/PhysRevLett.116.013001}

\bibitem{Gill2016}
Gill P 2016 {\em Journal of Physics: Conference Series\/} {\bf 723} 012053
  \urlprefix\url{http://stacks.iop.org/1742-6596/723/i=1/a=012053}

\bibitem{Bjerhammar1985}
Bjerhammar A 1985 {\em Bulletin g{\'e}od{\'e}sique\/} {\bf 59} 207--220 ISSN
  1432-1394 \urlprefix\url{http://dx.doi.org/10.1007/BF02520327}

\bibitem{Lisdat2016}
Lisdat C, Grosche G, Quintin N, Shi C, Raupach S, Grebing C, Nicolodi D,
  Stefani F, Al-Masoudi A, D\"orscher S, H\"afner S, Robyr J~L, Chiodo N,
  Bilicki S, Bookjans E, Koczwara A, Koke S, Kuhl A, Wiotte F, Meynadier F,
  Camisard E, Abgrall M, Lours M, Legero T, Schnatz H, Sterr U, Denker H,
  Chardonnet C, Le~Coq Y, Santarelli G, Amy-Klein A, Le~Targat R, Lodewyck J,
  Lopez O and Pottie P~E 2016 {\em Nature Communications\/} {\bf 7} 12443--
  \urlprefix\url{http://dx.doi.org/10.1038/ncomms12443}

\bibitem{Takano2016}
Takano T, Takamoto M, Ushijima I, Ohmae N, Akatsuka T, Yamaguchi A, Kuroishi Y,
  Munekane H, Miyahara B and Katori H 2016 {\em Nat Photon\/} {\bf 10} 662--666
  ISSN 1749-4885 \urlprefix\url{http://dx.doi.org/10.1038/nphoton.2016.159}

\bibitem{Chou2010}
Chou C~W, Hume D~B, Koelemeij J~C~J, Wineland D~J and Rosenband T 2010 {\em
  Phys. Rev. Lett.\/} {\bf 104}(7) 070802
  \urlprefix\url{http://link.aps.org/doi/10.1103/PhysRevLett.104.070802}

\bibitem{Inguscio2013}
Inguscio M and Fallani L 2013 {\em Atomic physics: precise measurements and
  ultracold matter\/} (Oxford University Press)

\bibitem{Godun2014}
Godun R~M, Nisbet-Jones P, Jones J, King S, Johnson L, Margolis H, Szymaniec K,
  Lea S, Bongs K and Gill P 2014 {\em Phys. Rev. Lett.\/} {\bf 113}(21) 210801
  \urlprefix\url{http://link.aps.org/doi/10.1103/PhysRevLett.113.210801}

\bibitem{Blatt2008}
Blatt S, Ludlow A~D, Campbell G~K, Thomsen J~W, Zelevinsky T, Boyd M~M, Ye J,
  Baillard X, Fouch\'e M, Le~Targat R, Brusch A, Lemonde P, Takamoto M, Hong
  F~L, Katori H and Flambaum V~V 2008 {\em Phys. Rev. Lett.\/} {\bf 100}(14)
  140801 \urlprefix\url{http://link.aps.org/doi/10.1103/PhysRevLett.100.140801}

\bibitem{Katori2003}
Katori H, Takamoto M, Pal'chikov V~G and Ovsiannikov V~D 2003 {\em Phys. Rev.
  Lett.\/} {\bf 91}(17) 173005
  \urlprefix\url{http://link.aps.org/doi/10.1103/PhysRevLett.91.173005}

\bibitem{Hinkley2013}
Hinkley N, Sherman J~A, Phillips N~B, Schioppo M, Lemke N~D, Beloy K, Pizzocaro
  M, Oates C~W and Ludlow A~D 2013 {\em Science\/} {\bf 341} 1215--1218
  \urlprefix\url{http://www.sciencemag.org/content/341/6151/1215.abstract}

\bibitem{Nemitz2016}
Nemitz N, Ohkubo T, Takamoto M, Ushijima I, Das M, Ohmae N and Katori H 2016
  {\em Nat Photon\/} {\bf 10} 258--261 ISSN 1749-4885
  \urlprefix\url{http://dx.doi.org/10.1038/nphoton.2016.20}

\bibitem{Akamatsu2014}
Akamatsu D, Yasuda M, Inaba H, Hosaka K, Tanabe T, Onae A and Hong F~L 2014
  {\em Opt. Express\/} {\bf 22} 7898--7905
  \urlprefix\url{http://www.opticsexpress.org/abstract.cfm?URI=oe-22-7-7898}

\bibitem{Park2013}
Park C~Y, Yu D~H, Lee W~K, Park S~E, Kim E~B, Lee S~K, Cho J~W, Yoon T~H, Mun
  J, Park S~J, Kwon T~Y and Lee S~B 2013 {\em Metrologia\/} {\bf 50} 119
  \urlprefix\url{http://stacks.iop.org/0026-1394/50/i=2/a=119}

\bibitem{Kohno2009}
Kohno T, Yasuda M, Hosaka K, Inaba H, Nakajima Y and Hong F~L 2009 {\em Applied
  Physics Express\/} {\bf 2} 072501
  \urlprefix\url{http://apex.jsap.jp/link?APEX/2/072501/}

\bibitem{Yasuda2012}
Yasuda M, Inaba H, Kohno T, Tanabe T, Nakajima Y, Hosaka K, Akamatsu D, Onae A,
  Suzuyama T, Amemiya M and Hong F~L 2012 {\em Applied Physics Express\/} {\bf
  5} 102401 \urlprefix\url{http://apex.jsap.jp/link?APEX/5/102401/}

\bibitem{Takamoto2015}
Takamoto M, Ushijima I, Das M, Nemitz N, Ohkubo T, Yamanaka K, Ohmae N, Takano
  T, Akatsuka T, Yamaguchi A and Katori H 2015 {\em Comptes Rendus Physique\/}
  {\bf 16} 489 -- 498 ISSN 1631-0705
  \urlprefix\url{http://www.sciencedirect.com/science/article/pii/S1631070515000730}

\bibitem{Porsev2004}
Porsev S~G, Derevianko A and Fortson E~N 2004 {\em Phys. Rev. A\/} {\bf 69}
  021403 \urlprefix\url{http://pra.aps.org/abstract/PRA/v69/i2/e021403}

\bibitem{Ludlow2011}
Ludlow A~D, Lemke N~D, Sherman J~A, Oates C~W, Qu\'em\'ener G, von Stecher J
  and Rey A~M 2011 {\em Phys. Rev. A\/} {\bf 84}(5) 052724
  \urlprefix\url{http://link.aps.org/doi/10.1103/PhysRevA.84.052724}

\bibitem{Pizzocaro2014}
Pizzocaro M, Calonico D, Pastor P~C, Catani J, Costanzo G~A, Levi F and Lorini
  L 2014 {\em Appl. Opt.\/} {\bf 53} 3388--3392
  \urlprefix\url{http://ao.osa.org/abstract.cfm?URI=ao-53-16-3388}

\bibitem{Pizzocaro2012}
Pizzocaro M, Costanzo G~A, Godone A, Levi F, Mura A, Zoppi M and Calonico D
  2012 {\em IEEE Trans. Ultrason., Ferroelect., Freq. Cont.\/} {\bf 59}
  426--431 ISSN 0885-3010

\bibitem{Pizzocaro2013}
Pizzocaro M, Calonico D, Calosso C, Clivati C, Costanzo G~A, Levi F and Mura A
  2013 {\em IEEE Trans. Ultrason., Ferroelect., Freq. Cont.\/} {\bf 60}
  273--280 ISSN 0885-3010

\bibitem{Blatt2009}
Blatt S, Thomsen J~W, Campbell G~K, Ludlow A~D, Swallows M~D, Martin M~J, Boyd
  M~M and Ye J 2009 {\em Phys. Rev. A\/} {\bf 80}(5) 052703
  \urlprefix\url{http://link.aps.org/doi/10.1103/PhysRevA.80.052703}

\bibitem{Katori2015}
Katori H, Ovsiannikov V~D, Marmo S~I and Palchikov V~G 2015 {\em Phys. Rev.
  A\/} {\bf 91}(5) 052503
  \urlprefix\url{http://link.aps.org/doi/10.1103/PhysRevA.91.052503}

\bibitem{Barber2006}
Barber Z~W, Hoyt C~W, Oates C~W, Hollberg L, Taichenachev A~V and Yudin V~I
  2006 {\em Phys. Rev. Lett.\/} {\bf 96} 083002

\bibitem{Targat2012}
Targat R~L, Lorini L, Gurov M, Zawada M, Gartman R, Nagórny B, Lemonde P and
  Lodewyck J 2012 Comparison of two strontium optical lattice clocks in
  agreement at the \num{e-16} level {\em European Frequency and Time Forum
  (EFTF), 2012\/} pp 19--22

\bibitem{Lemke2011}
Lemke N~D, von Stecher J, Sherman J~A, Rey A~M, Oates C~W and Ludlow A~D 2011
  {\em Phys. Rev. Lett.\/} {\bf 107}(10) 103902
  \urlprefix\url{http://link.aps.org/doi/10.1103/PhysRevLett.107.103902}

\bibitem{Campbell2009}
Campbell G~K, Boyd M~M, Thomsen J~W, Martin M~J, Blatt S, Swallows M~D,
  Nicholson T~L, Fortier T, Oates C~W, Diddams S~A, Lemke N~D, Naidon P,
  Julienne P, Ye J and Ludlow A~D 2009 {\em Science\/} {\bf 324} 360--363 ISSN
  0036-8075 \urlprefix\url{http://science.sciencemag.org/content/324/5925/360}

\bibitem{Sherman2012}
Sherman J~A, Lemke N~D, Hinkley N, Pizzocaro M, Fox R~W, Ludlow A~D and Oates
  C~W 2012 {\em Phys. Rev. Lett.\/} {\bf 108}(15) 153002
  \urlprefix\url{http://link.aps.org/doi/10.1103/PhysRevLett.108.153002}

\bibitem{Beloy2014}
Beloy K, Hinkley N, Phillips N~B, Sherman J~A, Schioppo M, Lehman J, Feldman A,
  Hanssen L~M, Oates C~W and Ludlow A~D 2014 {\em Phys. Rev. Lett.\/} {\bf
  113}(26) 260801
  \urlprefix\url{http://link.aps.org/doi/10.1103/PhysRevLett.113.260801}

\bibitem{BIPM2008a}
BIPM, IEC, IFCC, ILAC, ISO, IUPAC, IUPAP and OIML 2008 Evaluation of
  measurement data - guide to the expression of uncertainty in measurement
  Joint Committee for Guides in Metrology, JCGM 100
  \urlprefix\url{http://www.bipm.org/en/publications/guides/gum.html}

\bibitem{Margolis2013}
Margolis H, Godun R, Gill P, Johnson L, Shemar S, Whibberley P, Calonico D,
  Levi F, Lorini L, Pizzocaro M, Delva P, Bize S, Achkar J, Denker H, Timmen L,
  Voigt C, Falke S, Piester D, Lisdat C, Sterr U, Vogt S, Weyers S, Gersl J,
  Lindvall T and Merimaa M 2013 International timescales with optical clocks
  ({ITOC}) {\em European Frequency and Time Forum International Frequency
  Control Symposium (EFTF/IFC), 2013 Joint\/} pp 908--911

\bibitem{Denker2015a}
Denker H, Timmen L and Voigt C 2015 Gravity field modelling with regard to
  optical clock comparisons {\em XXVI General Assembly of the International
  Union of Geodesy and Geophysics (IUGG), Earth and Environmental Sciences for
  Future Generations, Prague, Czech Republic, 22 June--02 July 2015\/}
  \urlprefix\url{http://gghs2016.com/wp-content/uploads/2016/07/GGHS2016_paper_132.pdf}

\bibitem{Calonico2007}
Calonico D, Cina A, Bendea I~H, Levi F, Lorini L and Godone A 2007 {\em
  Metrologia\/} {\bf 44} L44
  \urlprefix\url{http://stacks.iop.org/0026-1394/44/i=5/a=N03}

\bibitem{Denker2015b}
Denker H, Timmen L and Voigt C 2015 private communication

\bibitem{Petit2010}
Petit G and Luzum B 2010 Iers conventions (2010) IERS Technical Note~36
  International Earth Rotation and Reference Systems Service (IERS) Frankfurt
  am Main iSBN 3-89888-989-6

\bibitem{Groten2004}
Groten E 2004 {\em Journal of Geodesy\/} {\bf 77} 724--797 ISSN 1432-1394
  \urlprefix\url{http://dx.doi.org/10.1007/s00190-003-0373-y}

\bibitem{Denker2013}
Denker H 2013 {\em Regional Gravity Field Modeling: Theory and Practical
  Results\/} (Berlin, Heidelberg: Springer) pp 185--291 ISBN 978-3-642-28000-9
  \urlprefix\url{http://dx.doi.org/10.1007/978-3-642-28000-9_5}

\bibitem{Lodewyck2012}
Lodewyck J, Zawada M, Lorini L, Gurov M and Lemonde P 2012 {\em IEEE Trans.
  Ultrason., Ferroelect., Freq. Cont.\/} {\bf 59} 411 --415 ISSN 0885-3010

\bibitem{Gibble2013}
Gibble K 2013 {\em Phys. Rev. Lett.\/} {\bf 110}(18) 180802
  \urlprefix\url{http://link.aps.org/doi/10.1103/PhysRevLett.110.180802}

\bibitem{Porsev2014}
Porsev S~G, Safronova M~S, Derevianko A and Clark C~W 2014 {\em Phys. Rev. A\/}
  {\bf 89}(1) 012711
  \urlprefix\url{http://link.aps.org/doi/10.1103/PhysRevA.89.012711}

\bibitem{Lemonde2005}
Lemonde P and Wolf P 2005 {\em Phys. Rev. A\/} {\bf 72}(3) 033409
  \urlprefix\url{http://link.aps.org/doi/10.1103/PhysRevA.72.033409}

\bibitem{Falke2012}
Falke S, Misera M, Sterr U and Lisdat C 2012 {\em Applied Physics B\/} {\bf
  107} 301--311 ISSN 1432-0649
  \urlprefix\url{http://dx.doi.org/10.1007/s00340-012-4952-6}

\bibitem{Luenberger1998}
Luenberger D~G 1998 {\em Optimization by Vector Space Methods\/} (New York:
  Wiley) ISBN 0471-18I17-X

\bibitem{Cox2006}
Cox M~G, Eiø C, Mana G and Pennecchi F 2006 {\em Metrologia\/} {\bf 43} S268
  \urlprefix\url{http://stacks.iop.org/0026-1394/43/i=4/a=S14}

\bibitem{Margolis2015}
Margolis H~S and Gill P 2015 {\em Metrologia\/} {\bf 52} 628
  \urlprefix\url{http://stacks.iop.org/0026-1394/52/i=5/a=628}

\bibitem{CIPM2015}
CIPM 2015 {Recommendation 2 (CI-2015): Updates to the list of standard
  frequencies}
  \urlprefix\url{http://www.bipm.org/jsp/en/CIPMRecommendations.jsp}

\end{thebibliography}

\end{document}